# Brain-wide functional imaging to highlight differences between the diurnal and nocturnal neuronal activity in zebrafish larvae

**Giuseppe de Vito**[†,1,2,3], **Lapo Turrini**[†,1,4], **Chiara Fornetto**[†,1,5], **Elena Trabalzini**[2], **Pietro Ricci**[1,6], **Duccio Fanelli**[4], **Francesco Vanzi**[1,7], **Francesco Saverio Pavone**[1,4,8,*]

[1]European Laboratory for Non-Linear Spectroscopy, Sesto Fiorentino, Italy.

[2]Department of Neuroscience, Psychology, Drug Research and Child Health, University of Florence, Florence, Italy.

[3]Interdepartmental centre for the study of complex dynamics, University of Florence, Sesto Fiorentino, Italy.

[4]Department of Physics and Astronomy, University of Florence, Sesto Fiorentino, Italy.

[5]Currently at: School of Life Sciences, University of Sussex, Brighton, UK.

[6]Currently at: Department of Applied Physics, Universitat de Barcelona, Barcelona, Spain.

[7]Department of Biology, University of Florence, Sesto Fiorentino, Italy.

[8]National Institute of Optics, National Research Council, Sesto Fiorentino, Italy.

[†]These authors contributed equally to this work.

**\* Correspondence:**
Corresponding Author
francesco.pavone@unifi.it

## Abstract

Most living organisms show highly conserved physiological changes following a 24-hour cycle which goes by the name of circadian rhythm. Among experimental models, the effects of light-dark cycle have been recently investigated in the larval zebrafish. Owing to its small size and transparency, this vertebrate enables optical access to the entire brain. Indeed, the combination of this organism with light-sheet imaging grants high spatio-temporal resolution volumetric recording of neuronal activity. This imaging technique, in its multiphoton variant, allows functional investigations without unwanted visual stimulation. Here, we employed a custom two-photon light-sheet microscope to study brain-wide differences in neuronal activity between diurnal and nocturnal periods in larval zebrafish assessed at the transition between day and night. We describe for the first time an activity increase in the low frequency domain of the pretectum and a frequency-localized activity decrease of the anterior rhombencephalic turning region during the nocturnal period. Moreover, our data confirm a nocturnal reduction in habenular activity. Furthermore, brain-wide detrended fluctuation analysis revealed a nocturnal decrease in the self-affinity of the neuronal signals in parts of the dorsal thalamus and the medulla oblongata and an increase in the pretectum.

Our data show that brain-wide nonlinear light-sheet imaging represents a useful tool to investigate circadian rhythm effects on neuronal activity.

## 1 Introduction

The majority of living organisms show metabolic, neural and behavioral changes following a 24-hour cycle which goes by the name of circadian rhythm (Huang et al., 2011). All of these manifestations are regulated by an endogenous molecular clock which is highly conserved through evolution (Dunlap and Loros, 2017). Indeed, earth's rotation around its axis and the consequent alternation of light and dark periods, produce predictable cyclic changes in the geophysical environment (e.g., light intensity, temperature, humidity, etc.) which have exposed living organisms to millions of years of constant selective pressure (Sharma, 2005). It is therefore not surprising that the capacity for biological timekeeping arose independently at least three times through evolution and that it is nearly ubiquitous among living organisms, from prokaryotes and plants, to fungi and animals (Dunlap and Loros, 2017). From the point of view of research, precisely this high grade of phylogenetic conservation allows for the use of model organisms to investigate the manifold implications of circadian rhythm in a translational perspective. Indeed, researchers have historically explored molecular and genetic aspects underlying circadian timekeeping abilities employing several organisms, ranging from cyanobacteria (Dvornyk et al., 2003; Cohen and Golden, 2015), to fungi of the *Neurospora* genus (Bell-Pedersen et al., 1996; Dunlap et al., 2007), *D. melanogaster* (Rosato et al., 2006; Dubowy and Sehgal, 2017) and mouse (Vitaterna et al., 1994; Takahashi et al., 2008). Conversely, the study of behaviourally relevant manifestations, and their alteration due to specific genetic mutations (Naylor et al., 2000; Yu and Weaver, 2011) or pharmacological treatments (Sulli et al., 2018; Tamai et al., 2018), has been typically conducted on mammal models.

Recently, another vertebrate, zebrafish (*Danio rerio*), has made its way into the panel of model organisms applied to the study of circadian rhythm. Indeed, this freshwater teleost has well-known distinctive features that can be applied to the study of previously inaccessible aspects related to circadian timekeeping, thus making this animal a valuable complement to mammalian models. The zebrafish genome shows a high grade of homology with the human genome, with the presence of orthologous genes involved in the control of the circadian clock such as *cry*, *clock*, *bmal* and *per* (Vatine et al., 2011), to cite just a few. This organism shows external fertilization and its larval form develops rapidly, responding to light stimuli with photomotor responses only 24 hours post fertilization (Kokel et al., 2010). Indeed, despite the larval zebrafish visual system starting to work from the third day of development (Easter and Nicola, 1996), due to the widespread expression of nonvisual opsins sensitive to UV/visible light (Davies et al., 2015; Steindal and Whitmore, 2020), zebrafish are entrained into light-dark cycle well before (Whitmore et al., 2000). In fact, zebrafish larvae are employed from the very first days of development in behavioral assays which, evaluating locomotor activity, aim at characterizing the behavioral manifestations of the circadian rhythm and those of its alterations (Cahill et al., 1998; Debruyne et al., 2004; Morbiato et al., 2019; Silva et al., 2022). Interestingly, zebrafish larvae are also employed for investigating the role of different brain districts and/or neurotransmitters in the regulation of physiological circadian oscillations (Elbaz et al., 2013; Oikonomou et al., 2019; Wang et al., 2020; Basnakova et al., 2021). Strikingly, owing to their small size, genetic tractability to generate transgenic lines expressing fluorescent neuronal activity reporters, and tissue transparency, zebrafish larvae are to date the only vertebrate model allowing whole-brain functional imaging at both high spatial and temporal resolution (Ahrens et al., 2013). This point is particularly critical since only employing this animal model we have the opportunity to investigate almost in real-time on a whole-organ level the diverse neuronal dynamics occurring in a vertebrate brain entrained into the circadian rhythm. Notably, a few recent studies have proficiently employed different imaging techniques to study the effects of circadian clock on larval zebrafish brain activity (Lee et al., 2017; Lin and Jesuthasan, 2017; Leung et al., 2019). Indeed, owing to the important advancements recently brought to fluorescence microscopy methods, and to light-sheet fluorescence microscopy (Stelzer et al., 2021; Ricci et al., 2022) amongst all, the exploration of the neuronal correlates underlying the circadian regulation of different brain states has become a novel and promising field of investigation. Recently, light-sheet fluorescence microscopy variants employing multiphoton excitation (Truong et al., 2011; Wolf et al., 2015; Keomanee-Dizon et al., 2020; Maioli et al., 2020) have emerged as a powerful tool to study neuronal circuitries in delicate physiological or pathological contexts. Indeed, owing to the use of near-infrared light as an excitation source (wavelengths scantily perceived by the majority of vertebrates, zebrafish included (Lewis, 1955; Jacobs, 2012)), non-linear light-sheet microscopy has been successfully applied in manifold applications ranging from the investigation of neuronal circuitries underlying phototaxis (Wolf et al., 2017) and numerosity capability (Messina et al., 2022) to the description of propagating seizure activity (de Vito et al., 2022), but not yet to the study of neuronal correlates underlying circadian brain state transitions.

In this work, we employed a custom two-photon light-sheet fluorescence microscope (2P LSFM) (de Vito et al., 2022) to investigate, with high spatio-temporal resolution, brain-wide neuronal dynamics of zebrafish larvae across the switch-over between light and dark (de Vito et al., 2020a) in order to highlight potential functional features distinctive of alertness and resting states.

## 2 Method

### 2.1 2P LSFM

The custom-made 2P LSFM was already described in detail in a previous article (de Vito et al., 2022), we will briefly recapitulate it here. The optical paths of the setup are shown in Supplementary Fig. S1.

The illumination beam at 930 nm is generated by a tunable Ti-Sa laser (Chameleon Ultra II, Coherent) and precompensated for the group delay dispersion with a pulse compressor (PreComp, Coherent). The laser power is attenuated exploiting a half-wave plate and a Glan-Thompson polarizer. The beam is scanned along the fronto-caudal direction of the larva by a resonant galvanometric mirror (CRS-8 kHz, Cambridge Technology) to generate the digital-scanned light-sheet. A second, closed-loop, galvanometric mirror (6215H, Cambridge Technology) is exploited to scan the light-sheet along the dorso-ventral direction of the larva at the frequency of 5 Hz.

The sample illumination is provided by two objectives (XLFLUOR4X/340/0,28, Olympus) placed at the two sides of the larva. An electro-optics modulator (84502050006, Qioptiq), a polarizing beam splitter and a total of two half-wave plates and a quarter-wave plate are used to alternatively switch the illumination between the two objectives at 100 kHz, while maintaining an horizontal polarization orientation on the sample. This arrangement grants a significant increase in signal generation and collection, as we showed in our previous articles (de Vito et al., 2020b, 2022).

A series of lenses placed between the galvanometric mirrors and the objectives magnifies the beam diameter 1.2 times, underfilling the objective pupils.

The zebrafish larva is placed inside a chamber filled with thermostated fish water. The chamber is laterally walled by glass surfaces (0.17 mm thickness) and is open on the top (where the detection objective is placed). The bottom of the chamber is made of light-diffusing (opaque) plexiglass.

A water-immersion objective is used to collect fluorescence signal (XLUMPLFLN20XW, Olympus, NA=1). The image is projected on a camera (ORCA-Flash 4.0 V3, Hamamatsu) by an optical system composed of two lenses and by a second objective (UPLFLN10X2, Olympus, NA=0.3). The final magnification on the camera sensor is 3×. Images were acquired at 16-bit depth.

An electrically-tunable lens (EL-16-40-TC-VIS-5D-C, Optotune), placed before the camera-side objective, is used to remotely scan the detection focal place, in order to synchronize its position with the light sheet without moving the objective.

Differently from the setup described in (de Vito et al., 2022), we used a scientific projector (DLP LightCrafter 4500, Texas Instruments) placed under the sample chamber, to illuminate the latter with red light (630 nm, FWHM 20 nm) through its transparent polycarbonate bottom. Moreover, we filtered the fluorescence light with a different (narrower) filter centered at 520 nm (FF01-520/60-25, Semrock). The infrared laser was blocked before the camera with a 750 shortpass filter (FF01-750/SP-25, Semrock).

## 2.2 Zebrafish larvae and imaging protocol

We observed 6 4-days-post-fertilization (dpf) transgenic larval zebrafish expressing GCaMP6s-H2B, a nuclear-localized pan-neuronal calcium indicator (Vladimirov et al., 2014), in homozygous albino background (genomic feature "b4", zfin ID: "ZDB-ALT-980203-365", affected genomic region: *slc45a2*, Vladimirov et al., 2014; Müllenbroich et al., 2018; ZFIN Feature: b4, n.d.). As previously reported (Whitmore et al., 2000; Wang et al., 2020), molecular and functional evidence demonstrates that at this stage of development zebrafish larvae are already entrained into the light-dark cycle. Fish maintenance and handling were carried out in accordance with European and Italian law on animal experimentation (Directive 2010/63/EU and D.L. 4 March 2014, n.26, respectively).

Larvae were entrained in an artificial 14/10-h light/dark cycle from their zygote stage until the conclusion of the experiment. Both during husbandry and during image acquisition, larvae were subjected to red-light illumination during artificial diurnal periods and complete darkness during artificial nocturnal periods. Preliminary behavioral recordings verified that no difference is observable between larvae exposed to white or red light during the diurnal phase of day-night cycle. For each larva, we recorded 5 minutes of neuronal calcium activity every 20 minutes for about 280 minutes (on average 7 acquisitions during the diurnal period and 7 acquisitions during the nocturnal period), covering the switch-over periods between day/night (sunset) or night/day (dawn) of the aforementioned artificial 14/10-h light/dark cycle. For half of the larvae we employed an artificial light/dark cycle protocol corresponding to the diurnal period at the beginning of the recording and the nocturnal period at the end of the recording ("forward transition", recording the transition at zeitgeber time 14); for the other half of the larvae we employed a protocol corresponding to the nocturnal period at the beginning of the recording and the diurnal period at the end of the recording ("reverse transition", recording the transition at zeitgeber time 0). Immediately before the acquisition the larvae were paralyzed with a myorelaxant agent (2mM, 10 min d-tubocurarine, 93750, Sigma-Aldrich, St. Louis, MO, USA), included in 1.5% low-melting-temperature agarose (Turrini et al., 2017) and mounted on a custom-made glass support immersed in fish water thermostated at 28.5 °C.
Imaging was performed at the volumetric frequency of 5 Hz, with a voxel size of 2 × 2 × 5 µm$^3$, and a field of view of about 1 × 1 × 0.15 mm$^3$. Cumulative laser power at the exit of the illumination objectives was set at 200 mW ± 5%, with an estimated power on the sample of 140 mW, as in Ref. (de Vito et al., 2022). This value was chosen after performing preliminary studies about phototoxicity described in the same article.

## 2.3 Image analysis

After the acquisition, the data were manually inspected and movement artifacts were removed, thus potentially generating multiple shorter temporal substacks from each original 5-minute stack.
Images were post-processed as described in Ref. (de Vito et al., 2022). We will recapitulate briefly the process here. First, by applying a binary mask to each stack we masked all pixels not belonging to the larva.
After masking, voxel-based 32-bit *ΔF/F$_0$* signals were computed using the following formula: $(F - F_0)/(F_0 - D)$, where $F_0$ is the voxel-based first decile value along the temporal dimension, while $D$ is the nearest lower integer of the lowest unmasked $F_0$ value.
Then, we spatially aligned brain volumes acquired from different animals in a post-processing passage, as we did in Ref. (de Vito et al., 2022). Briefly, time-lapse *ΔF/F$_0$* movies were first spatially interpolated along the axial dimension to an almost isotropic voxel size of 2.2 µm × 2.2 µm × 2.0 µm. Then, all stacks were aligned to an internal larval zebrafish reference brain. Finally, aligned stacks were binned (neglecting masked voxels) to the final voxel size of 4.4 µm × 4.4 µm × 4.0 µm. The alignment procedure was performed using a series of custom-made Python scripts made publicly available under the MIT License on GitHub ("https://github.com/lens-biophotonics/2P-LSFM-align").
To generate the power spectra, for each voxel we computed the square value of the modulus of the fast Fourier transform (FFT) of the *ΔF/F$_0$* signal (after the alignment procedure). Prior to the FFT computation, substacks were segmented in 1-minute portions (time-lapses shorter than 1 minute were discarded) to access a consistent frequency domain for further data analysis. Finally, a spatial bidimensional median filter (with kernel size: 3×3 pixels, limited to unmasked pixels) was applied for each plane and each frequency of the resulting power spectra. For Region-Of-Interest (ROI)-based analysis, power spectra of voxels belonging to the same ROI were averaged together. ROI-based standard deviation values were calculated by averaging the *ΔF/F$_0$* signal over each ROI for each substack.
For Detrended Fluctuation Analysis (DFA) (Peng et al., 1994), we employed the code developed by Dr. D. Krzemiński and available at: "https://github.com/dokato/dfa" under MIT license. Prior to α-value calculation, substacks were segmented in 1-minute portions (time-lapses shorter than 1 minute were discarded). Values of the detrended root mean square were computed on temporal windows ranging from 3.2 s to 58.8 s. For ROI-based analysis, α-values of voxels belonging to the same ROI were averaged.

In all cases, due to the effect of the motion artifact removal procedure and/or to the 1-minute segmentation, from an individual 5-minute recording window multiple substacks were generated. Data related to these substacks were averaged together in order to obtain an individual value for each 5-minute recording time-lapse. Only for the ROI-based standard deviation analysis the length of the substacks could differ (instead of being always equal to one minute). Therefore in this case a weighted mean was employed, with the numbers of original time frames as weights.

## 2.4 Statistical analysis

Differences between diurnal and nocturnal values were computed using general linear mixed models implemented in R (Kuznetsova et al., 2017). For these models, we used the period of the artificial day/night cycle (diurnal *vs* nocturnal) and the binarized recording time (before *vs* after the switch event) as fixed factors and the individual larva as a random factor.

For voxel-based analysis, models were implemented in a frequentist fashion with the "lmerTest" package. t-values and p-values of the diurnal/nocturnal comparisons were extracted directly from the models (from the corresponding fixed factor). q-values were obtained by correcting for multiplicity using the Benjamini and Hochberg procedure (Benjamini and Hochberg, 1995), with a false discovery rate of 0.1.

For ROI-based analysis and power spectrum comparisons, the aforementioned statistical model was implemented in a Bayesian fashion, with the "brms" package. In this case the statistical significance was assessed by computing the 95%-credibility interval of the factor estimate and then checking if it includes the zero value (i.e. no difference) or not.

Color maps were generated in the Hue, Saturation, Value (HSV) color space. t-values of the diurnal/nocturnal comparison or α-values were mapped on the hue channel and the statistical significance on the saturation channel. Value channel was generated from the global average of the fluorescence signal.

## 3 Results

Zebrafish larvae were placed in a 2P LSFM specifically devised to perform fast brain-wide imaging using infrared excitation light (de Vito et al., 2020a, 2020c, 2022; Turrini et al., 2022), as shown in Fig. 1A and Supplementary Fig. S1. Employing this wavelength for illumination allows us to record the neuronal activity without disrupting larval circadian rhythm (Emran et al., 2010). In addition, the microscope was equipped with a projector to provide red illumination for an artificial day-night cycle, thus enabling the investigation of neuronal activity both during diurnal and nocturnal phases. We performed brain-wide calcium imaging both during forward (day-night) and reverse (night-day) transitions (Fig. 1B).

We investigated the presence of possible differences between the diurnal and nocturnal neuronal activity in different regions of the brain. We first focused our analysis on the anterior rhombencephalic turning region (ARTR (Dunn et al., 2016), also termed hindbrain oscillator, HBO (Ahrens et al., 2013; Wolf et al., 2017)). Due to the oscillating nature of the nervous activity in this region (Ahrens et al., 2013), we deemed to use an analysis approach based on the frequency domain. Consequently, we computed the voxel-based power spectra of brain-wide recordings and investigated the presence of diurnal/nocturnal differences. From the voxel-based map, the presence of a difference in the power spectra of the rhombencephalic region is evident (Fig. 2A top-left). We manually counturned this region identified on the basis of functional differences in the nervous activity, thus defining a ROI for subsequent analysis. We then computed the ROI-averaged calcium traces (shown in Fig. 2B top, separately for right and left ARTR halves) and subjected them to further statistical analysis. In this way, we found a reduction in the calcium activity in the ARTR ROI during the nocturnal period that resulted statistically significant (zero difference between the diurnal and nocturnal periods not included in the 95% credibility interval of the estimate) around the 0.12 Hz frequency (Fig. 2A bottom-left). In addition, we also compared the global activity in this ROI between the two periods by confronting the standard deviation values (Supplementary Fig. S2). However in this case the reduction in nervous activity associated with the nocturnal period ($-5.14*10^{-3}$ reduction, credibility interval: $[-12.65; 2.37]*10^{-3}$) was not found statistically significant, probably because the effect around the 0.12 Hz frequency was diluted by the reduced difference at the other frequencies. Using the voxel-based map generated by the difference in the nocturnal and diurnal power spectra, we also observed another effect of the nocturnal period, this time restricted to pretectum and associated with an increase during the nocturnal activity localized in the slowest frequency domain (Fig. 2A top-right). Also in this case, we first manually traced a ROI based on this difference in functional activity and then we further analyzed the ROI-averaged calcium traces (shown in Fig. 2B bottom). We found a statistically significant (significance as described above and in *Method*) increase in nervous activity for frequencies included between 0.02 Hz e 0.15 Hz (Fig. 2A bottom-right). In addition, also in this case we compared the global activity in this ROI between the two periods by confronting the standard deviation values. For this ROI we found a statistically significant increase ($7.70*10^{-3}$ increase, credibility interval: $[2.97; 12.42]*10^{-3}$) in the nervous activity associated with the nocturnal period (Fig. 2C).

Moreover, we found a reduction in the neuronal activity of the habenulae during the nocturnal phase, as shown in Supplementary Fig. S3 ($-7.47*10^{-3}$ reduction in the standard deviation of the ROI-averaged calcium traces, credibility interval: $[-12.29; -2.50]*10^{-3}$). This result confirms what recently reported by Basnakova and colleagues (Basnakova et al., 2021).

We then investigated the degree of self-affinity of the calcium signal exploiting DFA (Peng et al., 1994; Hardstone et al., 2012) to extract the scaling exponent (Mandelbrot and Wallis, 1969) ($\alpha$-value). A maximal value of self-affinity means that the process is a statistical fractal and it is defined by the absence of a characteristic temporal scale. If the degree of self-affinity is high but not maximal (as it is quite frequent with biological systems), then the signal scale-free properties bear a partial resemblance to that of a perfect statistical fractal. In this context, a high degree of self-affinity means that rescaled versions of small segments of the time series of a non-stationary stochastic process (such as the calcium trace) have similar standard deviation values as larger parts (Hardstone et al., 2012). We observed an $\alpha$-value greater than 0.5 in neuron-rich areas of the brain (Fig. 3A), indicating that the voxel-based calcium traces are characterized by positive autocorrelation. We proceeded to compare the voxel-based $\alpha$-values between the nocturnal and diurnal periods. We found two areas where the $\alpha$-values are lower during the nocturnal period: part of the medulla oblongata (MO, Fig. 3B left) and part of the dorsal thalamus (DT, Fig. 3B right).

We then computed the ROI-averaged calcium traces to subject them to further statistical analysis. Representative calcium traces from MO and DT ROIs are shown in Fig. 3C. We compared the ROI-averaged α-values between the circadian phases for both the ROIs (Fig. 3D) and we confirmed a significant decrease in these values in the nocturnal period (-0.038 decrease, credibility interval: [-0.058; -0.018] for DT and -0.050 decrease, credibility interval: [-0.071; -0.030] for MO). Moreover, we also analyzed the ROIs initially identified in the frequency-based approach to check for differences in the α-values. We did not find significant differences for the ARTR and the habenulae ROIs, however we observed an increase of the α-value for the pretectum ROI during the night (Fig. S4, 0.030 increase, credibility interval: [0.003; 0.057]).

Finally, by checking that the points on the fluctuation-versus-window size logarithmic plot lay on approximate straight lines across a wide range of window sizes (for all the voxels pertaining to the MO and DT ROIs), we verified that the α-values correctly reflect the degree of auto-correlation (or anticorrelation) of the signals (Supplementary Fig. S5).

## 4    Discussion

We performed brain-wide functional imaging to study the differences in neuronal activity between the diurnal and nocturnal periods in 4 dpf zebrafish larvae. We took great care not to affect the circadian rhythm of the animals. In particular, we employed infrared light for imaging excitation and a projector operated in the red spectrum to maintain larvae entrained in the circadian cycle even during measurements. Moreover, by switching the observation periods (day-night and night-day shifts), we controlled for the effects induced by the prolonged imaging sessions.

Our data confirm the nocturnal reduction in nervous activity in the habenulae previously observed by Basnakova and colleagues (Basnakova et al., 2021). Moreover, we describe for the first time a decrease in nervous activity in the ARTR/HBO (around 0.12 Hz). A phasic activation of the ARTR/HBO region was reported in the literature to precede the larva turning movements (Dunn et al., 2016; Wolf et al., 2017). Consequently, the effect that we observed in this region could be related with the known reduction in locomotor activity during the nocturnal period (Cahill et al., 1998; Hurd et al., 1998). We also describe for the first time an increase in the low frequency domain (from 0.02 to 0.15 Hz) of the activity in the pretectum during the nocturnal period. The pretectum was described in mammals to act, together with the superior colliculus (homologous of the optic tectum in zebrafish), on the tuning of the sleep/wake cycle to the environmental illumination (Miller et al., 1998; Zhang et al., 2019). Furthermore, the observed difference in this region resembles the reported activity in the infra-slow frequency band in dorsal pallium observed by Leung and collaborators during a slow-wave-sleep analogous in 7-14 dpf larvae (Leung et al., 2019).

Applying the DFA on imaging data of an entire vertebrate brain, we observed that neuron-rich areas of the brain are associated with α-values greater than 0.5. This means that the presence of larger signal fluctuations on longer time-scales is more probable than what could be expected only by chance (Hardstone et al., 2012). This result was actually expected, because this situation is reminiscent of pink noise (α-value≈1): a condition that is frequently encountered in biological systems and is considered a signature of the intrinsic dynamic complexity of brain activity (Gilden, 2001; Van Orden et al., 2003). We observed a reduction of the α-values in a region of the DT and in a region of the MO in the nocturnal period, indicating a decrease in the self-affinity of the neuronal signals. This unexpected result means that the pink-noise condition is partially lost during the nocturnal period in some regions of the brain. Parallelly, we found an α-value increase for the pretectum. This increase should be expected, since the observed increase in the low frequencies of the power spectrum is linked with the appearance of slow and large calcium oscillations such the ones shown in Fig. 2B (bottom row). This kind of oscillations are ostensibly at the basis of a growth in the autocorrelation of the signal.

We shall point out that our study also presents some limitations. The first being that the hypothalamus, a region heavily involved in sleep and responsabile of the hypocretin neuropeptide precursor production, owing to its depth into the brain was not imaged. Moreover, we opted to ignore the regions at anatomical borders in our analysis. This is because these regions are the most sensitive to movement artifacts and, even though the affected frames were manually identified and removed, some uncontrolled effects could still be present. In addition, we observed a progressive reduction of the nervous activity as the imaging time progressed. This effect was experimentally compensated by the recording protocol that we adopted (recording both the "forward" and "reverse" transitions) and statistically compensated by including the binarized recording time as a fixed factor in our statistical model. However, for these reasons, we deemed to focalize our ROI analysis only to the regions where the nocturnal/diurnal effect is stronger, as evidenced by the preliminary voxel analysis. We believe that our work will be useful for the community by identifying the most promising brain regions to focus future research in this field. At the same time, we cannot exclude the presence of significant differences in other brain regions, and we hope that our work will stimulate future research for a thorough and systematic study based on anatomically-defined ROIs.

In conclusion, we deem that the use of brain-wide functional imaging employing non-visible excitation light will be greatly beneficial to further progress the knowledge of the circadian rhythm effects on the neuronal activity of the zebrafish larvae, with important translational perspectives.

## 5    Conflict of Interest

The authors declare that the research was conducted in the absence of any commercial or financial relationships that could be construed as a potential conflict of interest.

## 6    Author Contributions

GdV and CF performed whole-brain imaging experiments; CF and FV conceived the study; GdV, ET, LT, CF and PR performed data processing; GdV performed data analysis; DF, FV and FSP provided critical analytical expertise; GdV and LT cured data presentation and wrote the manuscript. All authors have read and agreed to the published version of the manuscript.

## 7    Funding

This research was supported by the European Research Council (ERC) under the European Union's Horizon 2020 Research and Innovation program (grants agreement n.692943 and n.966623). This research has also been supported by the Italian Ministry for Education, University, and Research in the framework of the Advanced Lightsheet Microscopy Italian Node of Euro-Bioimaging ERIC, and by Bank Foundation Fondazione Cassa di Risparmio di Firenze with grant 2020.1991 "Human Brain Optical Mapping".

## 8    Acknowledgments

We thank Dr. Dominik Krzemiński from University of Cambridge for the publicly available Python algorithm to compute DFA. We thank Dr. Alberto Ferrari from Altoida Inc. for the useful discussion about the employed statistical analysis.

**Figures**

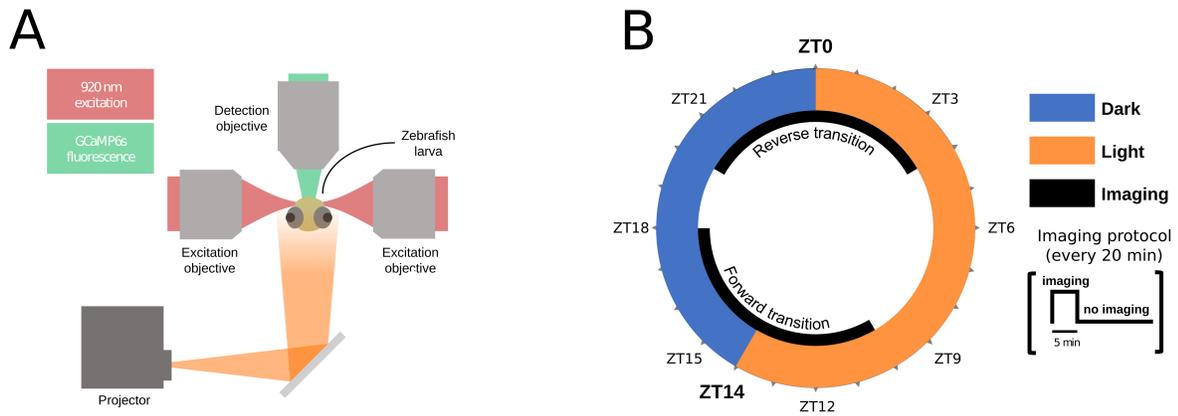

**Figure 1. (A)** Schematic representation of the experimental setup used for recording the larva neuronal activity. Rostral view. **(B)** Schematic representation of the experimental protocol. ZT: Zeitgeber Time. Half of the larvae were recorded at the ZT0 transition and the other half at the ZT14 transition.

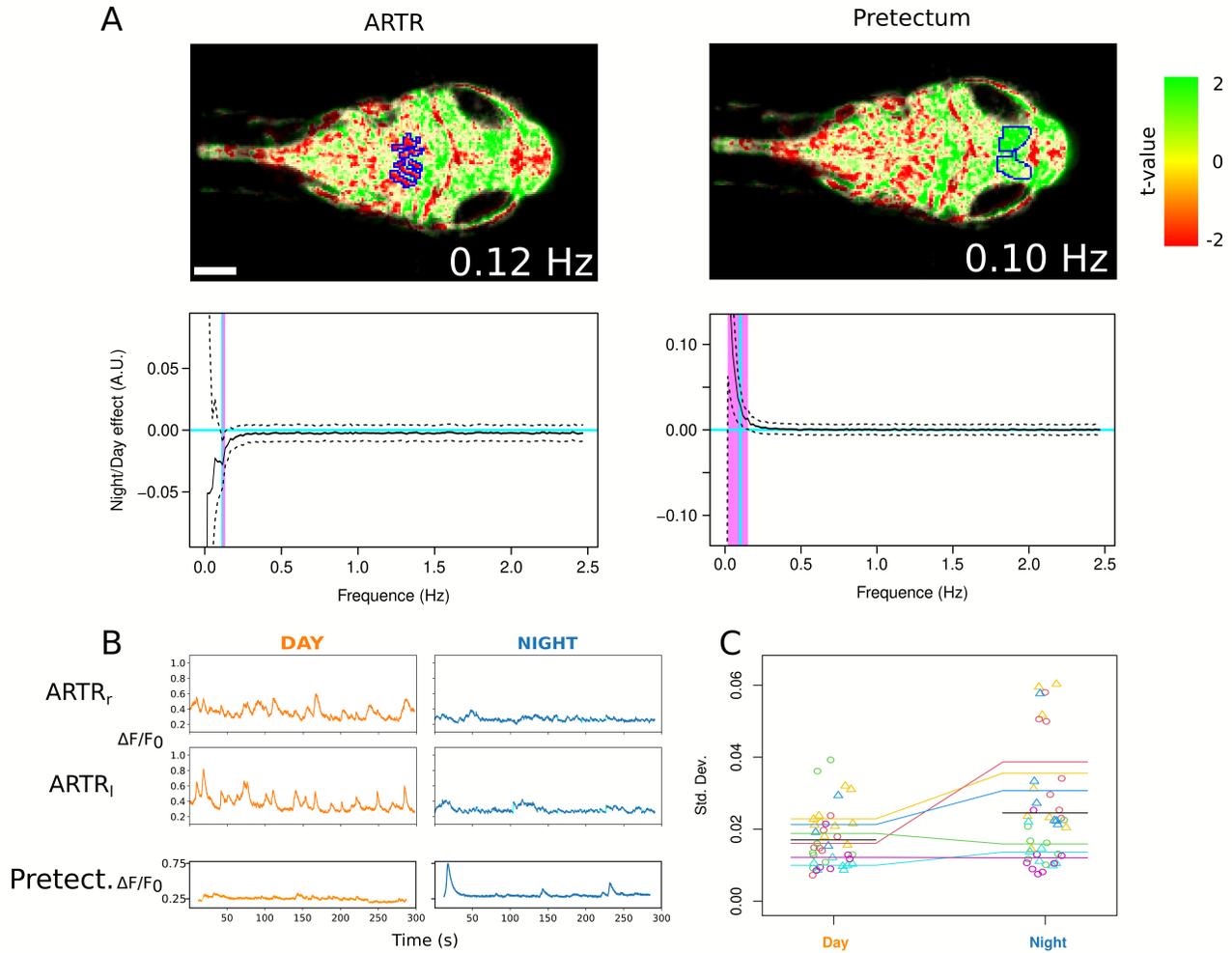

**Figure 2. (A)** Top: color maps of the larval brain (transversal sections, aggregate results from all the larvae) depicting the t-values of the effect of the nocturnal period on the frequency component at 0.12 Hz (left) and at 0.10 Hz (right) of the power spectrum of calcium activity. Hues as in the color bar on the right, saturation is maximal only for voxels displaying statistical significance. Blue outlines indicate the ARTR (left) and the pretectum (right) ROIs. Scale bar: 100 µm. Bottom: continuous black lines represent ROI-averaged statistical effect of the nocturnal period on the power spectra of the calcium activity of ARTR (left) and pretectum (right) ROIs. Dashed black lines represent credibility intervals at 95%. Vertical magenta bars indicate statistical significance. Horizontal cyan lines mark zero and vertical cyan lines mark 0.12 Hz (left) and 0.10 Hz (right). **(B)** Representative calcium traces (ROI-averaged) from the right and left ARTR halves (top, from one fish) and from the pretectum (bottom, from a second fish). Left: diurnal period; Right: nocturnal period. Lighter colors indicate segments missing due to motion artefacts. **(C)** Standard deviation values (one point for each individual recording) of average calcium traces of the pretectum in the diurnal and nocturnal periods. Different colors indicate different animals, circles indicate data acquired during the "forward" transition and triangles indicate data acquired during the "reverse" transition. Colored lines indicate the average values for each animal. Black horizontal segments indicate the diurnal and nocturnal global means.

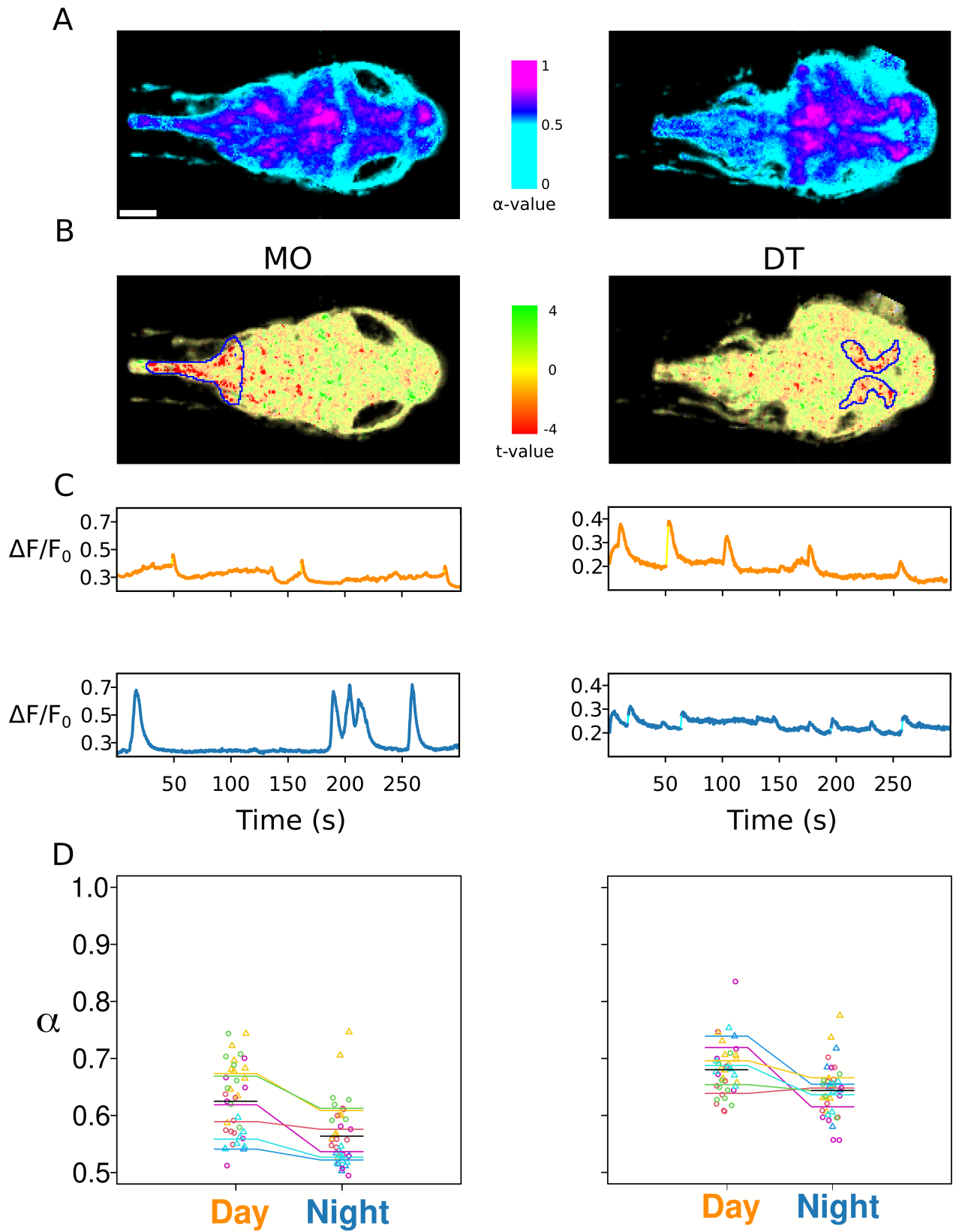

**Figure 3. (A)** Color maps of the larval brain (two transversal sections with the one on the right being 48 micron more ventral than the one on the left, aggregate results from all the larvae) depicting the average α-values of the calcium activity. Hue as in the color bar. Scale bar: 100 μm. **(B)** Color maps of the larval brain (transversal sections, aggregate results from all the larvae, section depths as in (A)) depicting the t-values of the effect of the nocturnal period on the α-values of the calcium activity. Hue as in the color bar, saturation is maximal only for voxels showing statistical significance. Blue outlines indicate MO (left) and DT (right) ROIs. Scale as in (A). **(C)** Representative calcium traces (ROI-averaged) from the MO (left) and from the DT (right) ROIs. Orange: diurnal period; blue: nocturnal period. Lighter colors indicate segments missing due to motion artifacts. **(D)** Average α-values (one point for each individual recording) of the MO (left) and DT (right) ROIs in the diurnal and nocturnal periods. Different colors indicate different animals, circles indicate data acquired during the "forward" transition and triangles indicate data acquired during the "reverse" transition. Colored lines indicate the average values for each animal. Black horizontal segments indicate the diurnal and nocturnal global means.

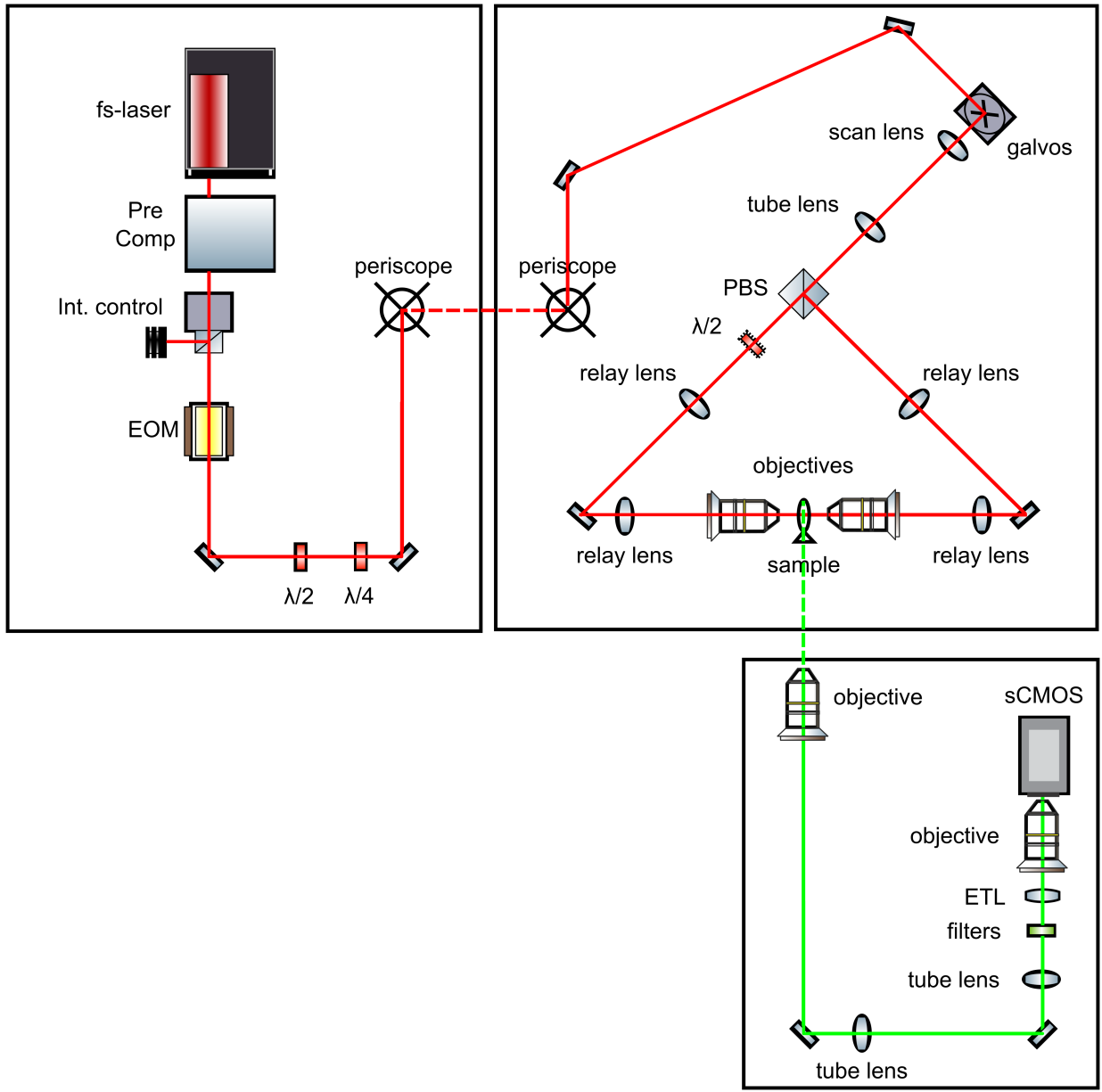

**Supplementary Figure S1.** Schematic of the custom-made 2P LSFM. fs-laser: femtosecond laser. Pre Comp: group delay dispersion precompensation unit. Int. control: power intensity control unit. EOM: electro-optic modulator. λ/2: half-wave plate. λ/4: quarter-wave plate. galvos: hybrid galvanometric mirrors assembly. PBS: polarizing beam splitter. ETL: electrically-tunable lens. Red lines: excitation beampath, green lines: detection beampath. Dashed segments indicate vertical paths. Figure and caption reproduced with permission from "Giuseppe de Vito, Lapo Turrini, Caroline Müllenbroich, Pietro Ricci, Giuseppe Sancataldo, Giacomo Mazzamuto, Natascia Tiso, Leonardo Sacconi, Duccio Fanelli, Ludovico Silvestri, Francesco Vanzi, and Francesco Saverio Pavone, "Fast whole-brain imaging of seizures in zebrafish larvae by two-photon light-sheet microscopy," Biomedical Optics Express 13, 1516-1536 (2022), doi: 10.1364/BOE.434146", © 2022 Optica Publishing Group.

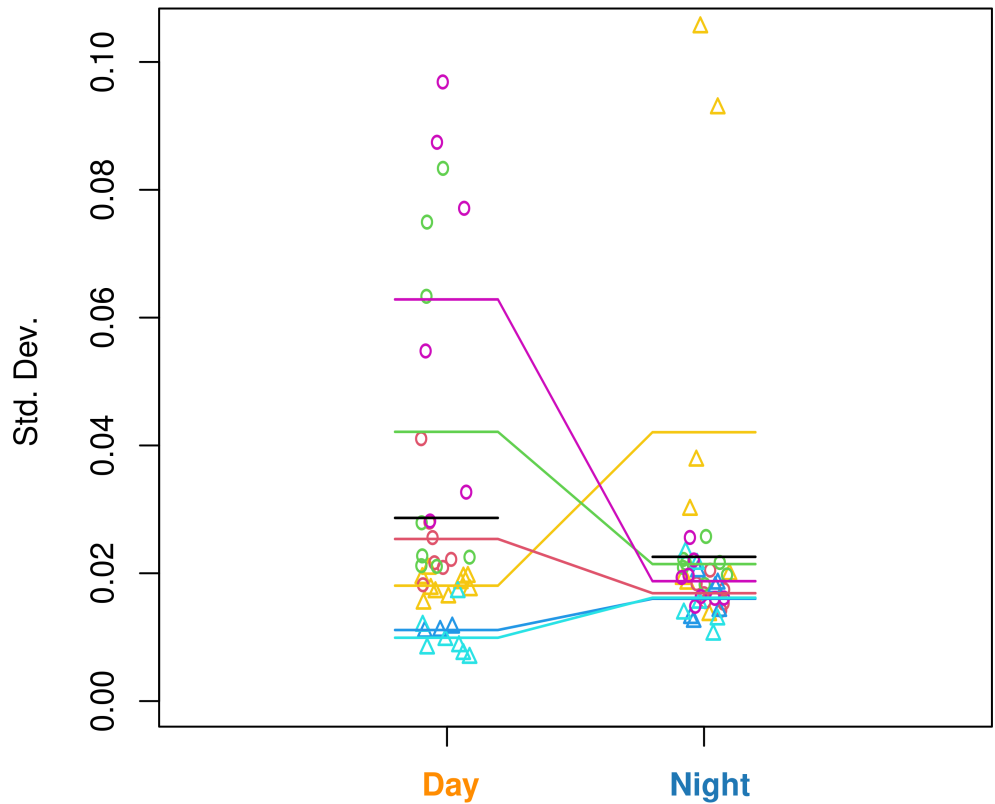

**Supplementary Figure S2.** Standard deviation values (one point for each individual recording) of average calcium traces of the ARTR ROI in the diurnal and nocturnal periods. Different colors indicate different animals, circles indicate data acquired during the "forward" transition and triangles indicate data acquired during the "reverse" transition (see "*Method*"). Colored lines indicate the average values for each animal. Black horizontal segments indicate the diurnal and nocturnal global means.

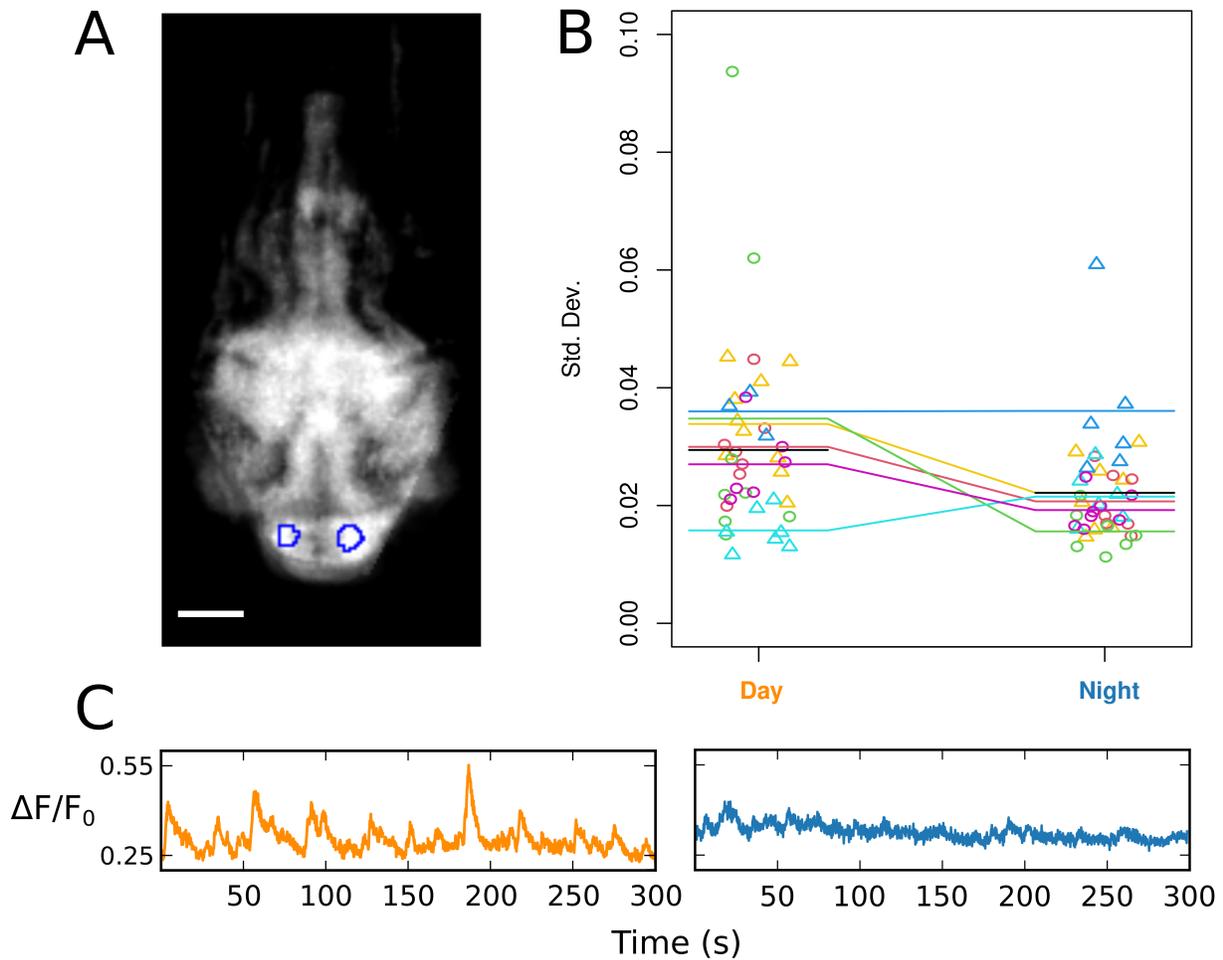

**Supplementary Figure S3. (A)** Transversal section of the larval brain obtained averaging all the animals. Blue outlines indicate the habenula ROI. Scale bar: 100 μm. **(B)** Standard deviation values (one point for each individual recording) of average calcium traces of the habenula ROI in the diurnal and nocturnal periods. Different colors indicate different animals, circles indicate data acquired during the "forward" transition and triangles indicate data acquired during the "reverse" transition (see "*Method*"). Colored lines indicate the average values for each animal. Black horizontal segments indicate the diurnal and nocturnal global means. **(C)** Representative calcium traces (ROI-averaged) from the habenula ROI. Orange: diurnal period; blue: nocturnal period.

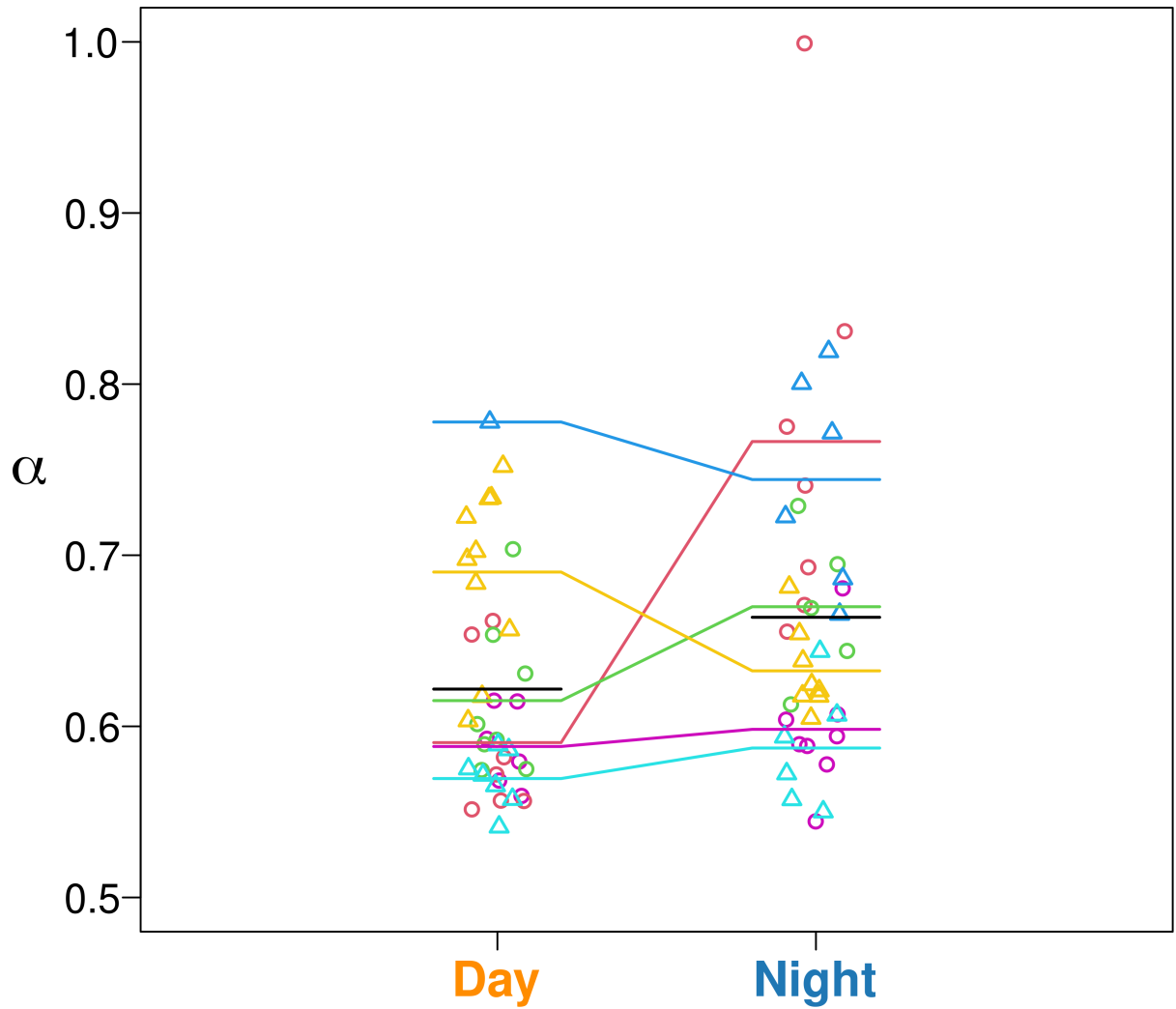

**Supplementary Figure S4.** Average α-values (one point for each individual recording) of the pretectum ROI in the diurnal and nocturnal periods. Different colors indicate different animals, circles indicate data acquired during the "forward" transition and triangles indicate data acquired during the "reverse" transition. Colored lines indicate the average values for each animal. Black horizontal segments indicate the diurnal and nocturnal global means.

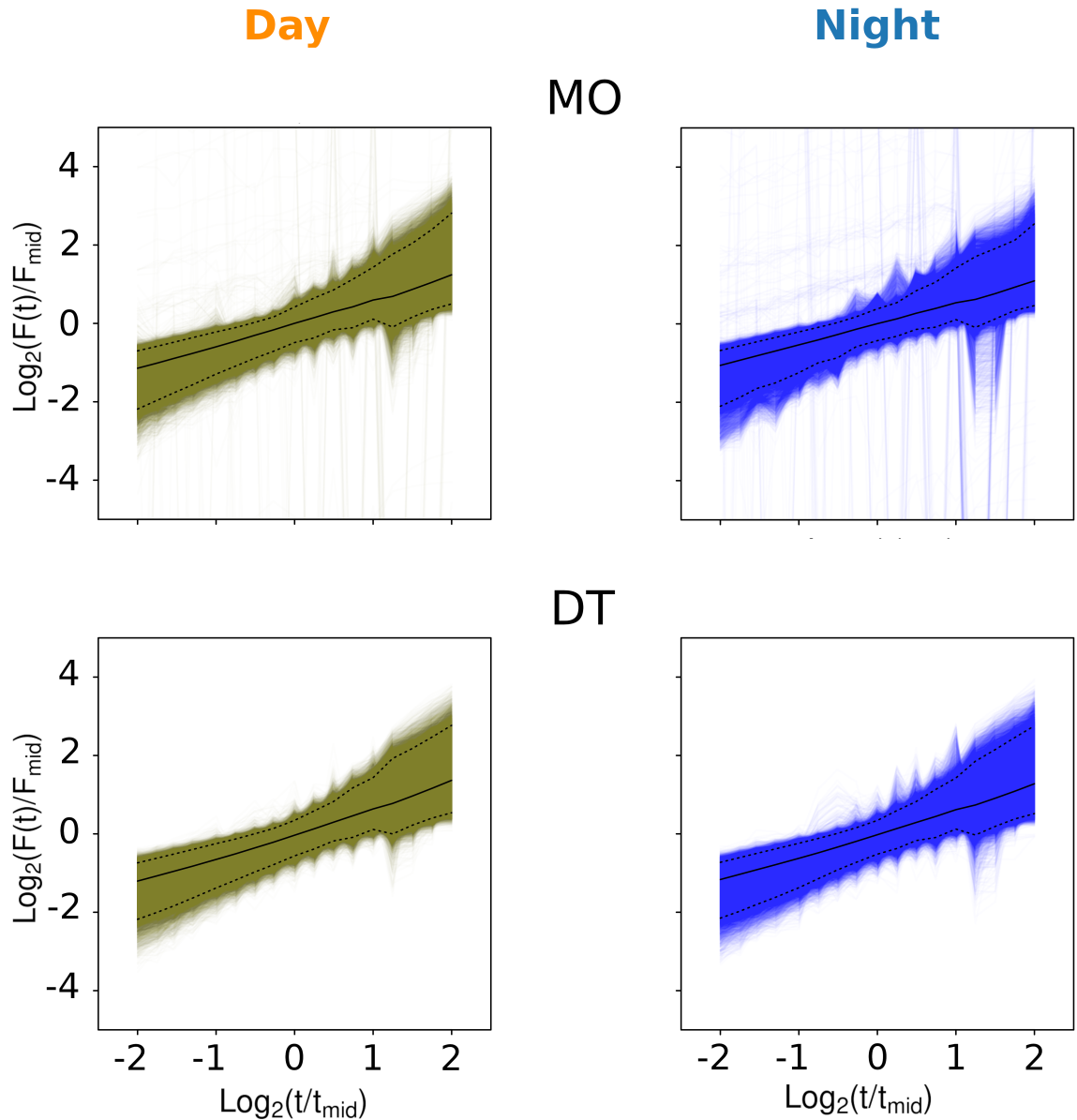

**Supplementary Figure S5.** Logarithmic plots of the mean signal fluctuations as a function of the time window lengths for all the voxels belonging to the MO (top) and DT (bottom) ROIs for the diurnal (left) and nocturnal (right) periods. Continuous black lines indicate the median values and dashed lines indicate the 2.5th and 97.5th percentiles. Individual colored lines connect all points belonging to the same voxel.